# Transient grating spectroscopy: An ultrarapid, nondestructive materials evaluation technique

Felix Hofmann*, Michael P. Short*, Cody A. Dennett

Structure-property relationships are the foundation of materials science. Linking microstructure and material properties is essential for predicting material response to driving forces, managing in-service material degradation, and engineering materials for optimal performance. Elastic, thermal, and acoustic properties provide a convenient gateway to directly or indirectly probe material structure across multiple length scales. We review how using the laser-induced transient grating spectroscopy (TGS) technique, which uses a transient diffraction grating to generate surface acoustic waves (SAWs) and temperature gratings on a material surface, non-destructively reveals the material's elasticity, thermal diffusivity, and energy dissipation on the sub-microsecond timescale, within a tunable sub-surface depth. This technique has already been applied to many challenging problems in materials characterization, from analysis of radiation damage, to colloidal crystals, to phonon-mediated thermal transport in nanostructured systems, to crystal orientation and lattice parameter determination. Examples of these applications, as well as inferring aspects of microstructural evolution, illustrate the wide potential reach of TGS to solve old materials challenges, and to uncover new science. We conclude by looking ahead at the tremendous potential of TGS for materials discovery and optimization when applied *in situ* to dynamically evolving systems.

**Introduction:**

Uncovering and exploiting structure-property relationships has been the core of the science and application of materials since antiquity. Almost five millenia ago, it was discovered that forging – controlling the microstructure of a piece of metal – imparted beneficial properties in addition to its composition. Since then the introduction and control of defects in materials, ranging across all classes and applications, has been central to materials optimization. From precise dopants in semiconductors (*1*), to cold or hot working metals (*2*), to doping of ceramics to boost electrical conductivity (*3*), to controlling crystallinity and polymer chemistry in plastics (*4*), the precise knowledge of material defects and our ability to control them is behind almost every material-based innovation from past to present.

Our ability to create and characterize materials struggles to keep pace with the application-driven need for innovation. Recent advances such as combinatorial thin film synthesis (*5*) and computational materials discovery (*6*) have dramatically accelerated our ability to design and synthesize new materials. Novel *in situ* techniques, such as micropillar compression (*7, 8*), liquid cell transmission electron microscopy (TEM), scanning electron microscopy (SEM) (*9*), and X-ray diffraction and imaging techniques (*10–12*) rapidly uncover dynamic material changes in response to driving forces such as stress, temperature, or changing chemistry. However, gaps in our ability to rapidly innovate still remain in some fields of study.

The perhaps oldest field of materials science – metallurgy – is a prime example where material innovation and design is limited by the speed at which characterization may be carried out, in part due to the lack of suitable characterization techniques across time and length scales. New alloys are often "discovered" by combinatorial trial-and-error, resulting in relatively few classes of high-performance alloys that are tailored for certain applications. In energy applications these alloys must withstand aggressive chemistries, high temperatures, and high pressures for extraordinarily long periods of time – decades for typical fossil and nuclear power plant applications. Coupled with the glacial timescale of licensing and regulation for mission-critical materials (*13*), this puts a serious bottleneck on the speed of materials innovation.

Fields like radiation materials science suffer even more, as the time, cost, and associated safety issues endemic to any radiation experiment render them among the slowest to produce useful data. This is largely due to the slow pace at which suitable materials degradation information, either direct or indirect, can be generated, which is itself limited by the slow speed, geometrical specimen constraints, and/or cost of undertaking these analyses. As an illustrative example, industrial nuclear reactor pressure vessels (RPVs) incorporate a large number of "inspection coupons" (*14*) made from the same material and positioned inside the vessel. Destructive mechanical tests, such as Charpy impact, (*15*) are used to monitor ductile-to-brittle transition temperature (DBTT) shifts as a quantitative measure of RPV health. However, many commercial US power plants have now run out of inspection coupons, leading operators to re-insert older coupons without full knowledge of the operating history, or they must remove material from the bulk RPV for testing (*16*, *17*).

Even scientific discovery in radiation materials science is fraught with empiricism, a scarcity of experiments, and often under-sampled data due to the time, cost, and difficulty associated with conducting experiments. For example, a recent study of the void swelling of three high-performance, ferritic/martensitic steels was conducted to see when void swelling would begin (*18*), signaling the rapid reduction in ductility of these materials (*19*). After almost a year of work, just six to eight radiation doses per material were probed (see Figure 1). What was ultimately sought was the *incubation dose* of void swelling which, if known, would set the useful lifetime of these high-performance materials. To scope out just three materials in a year simply is not fast enough to gather data in time for reactor licensing and deployment.

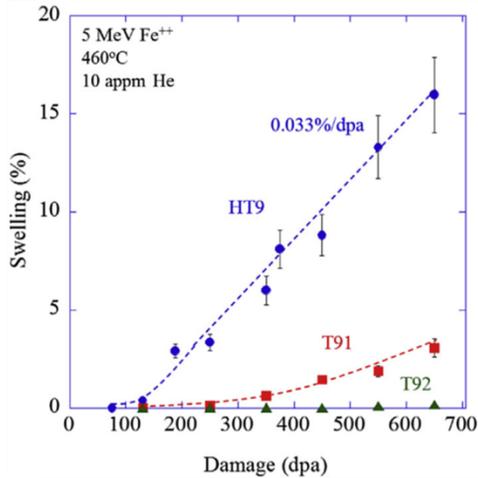 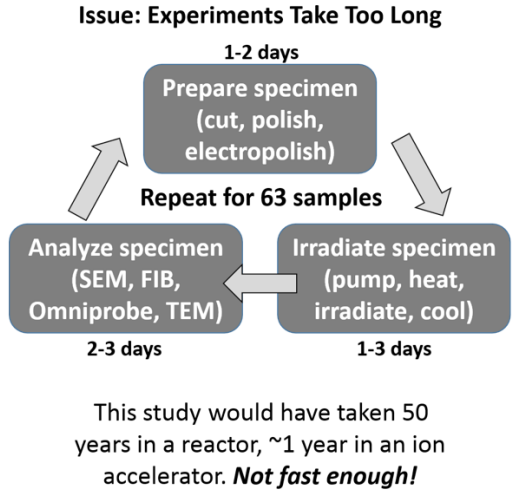

*Figure 1: Example of a typical radiation materials science experiment (18), illustrating the time required to obtain the desired data and its relative scarcity. Left hand figure reprinted from Getto et al. (18) with permission from Elsevier.*

These examples motivate the development and usage of far more rapid, non-destructive techniques. The desire is to use such methods to monitor microstructural evolution during heat treatment, mechanical testing, or even as *in situ* characterization in mission-critical applications across a wide range of fields; from nuclear to aerospace to fossil and advanced energy. However, with the regulator-required data consisting of mechanical properties, corrosion resistance, fatigue performance, and in some cases radiation damage resistance, it is often the case that laboratory experiments simply cannot produce the required data without actually building the reactor, resulting in a Catch 22 situation. A technique which can rapidly and non-destructively probe microstructural evolution and material properties, whether directly or indirectly, is the key to expediting innovation in these materials science fields just to keep pace with the speed of idea generation and the demands of the market (*18*).

This need for rapid *in situ* property determination is not limited to materials for present and future nuclear power, but spans a wide range of applications from the development of phononic metamaterials (*20, 21*), to clarifying the properties of glass forming liquids (*22, 23*), exploring the dynamics of granular materials (*24, 25*), developing optimized thermoelectrics (*26*), or designing metal-organic frameworks for sensing and catalysis (*27–29*) to name but a few examples. The so-called transient grating spectroscopy (TGS) technique is a versatile tool that goes some way towards addressing this need by providing a convenient, non-contact method of characterizing elastic, thermal, and acoustic properties at the micro-scale on microsecond time-scales. By pre-determining the relationship between these properties and other quantities of interest, TGS further offers the potential to act as an NDE technique even for field-inspection (one of the first commercial applications of TGS was thin film metrology (*30*)). Importantly TGS works in air, vacuum, or in any transparent fluid, at low or high temperatures, and in many *in operando*-like service conditions.

This review begins with an overview of the TGS technique, considering signal generation, interpretation, and practical implementation of TGS measurements. Recent applications of TGS to studying material property evolution will then be reviewed, focusing in particular on irradiation induced changes. Finally, an outlook is provided, highlighting the tremendous opportunities afforded by *in situ* TGS measurements for data-rich characterization of dynamically evolving systems.

**Transient Grating Spectroscopy - The Technique:**

TGS works by generating surface acoustic waves (SAWs) and a well-defined temperature grating. Two short excitation laser beams (typically <500 ps pulse duration, ~500 nm wavelength, ~ few μJ energy) are overlapped on the sample surface with a fixed angle $\theta$ (Fig. 2(a)). Interference of the beams generates a spatially periodic intensity pattern with wavelength $L$ on the sample surface (*31, 32*):

$$L = \frac{\lambda}{2 \sin(\frac{\theta}{2})}, \tag{1}$$

where $\lambda$ is the wavelength of the excitation light. Interference of excitation beams with the same polarization leads to the formation of a periodic intensity pattern, or, using excitation beams with crossed polarization, it is also possible to form a grating with a periodic polarization pattern (linear, elliptical, and circular) (*33*). Absorption of the light within the intensity grating leads to a spatially periodic temperature profile in the sample, which, due to thermal expansion, causes a spatially periodic variation of the specimen surface height. In addition, rapid thermal expansion launches two monochromatic, counter-propagating SAWs with wavelength $L$.

The SAWs and temperature grating decay can be monitored using a probe laser beam that is diffracted from the transient gratings at the sample surface. The intensity of the diffracted probe beam is spatially overlapped with a reference beam reflected from the sample surface and used in a heterodyne amplification scheme (*31, 32*). The intensity in the geometric channel containing both the diffracted signal and the reflected reference beam is monitored as a function of time after excitation using a fast avalanche photodiode and oscilloscope (Fig. 2(b)). By averaging over a few thousand exposures, low noise time traces can be generated in a few seconds per point (*34*). As an example, Fig. 2(c) shows the phase grating signal measured from a tungsten sample (blue trace) (*35*). Two components of this signal are clear: 1. an underlying decay, due to the thermal equilibration of the induced transient gratings in the sample (green dashed line). 2. oscillations due to the counter-propagating SAWs. Also shown is a trace from a region of the sample that was implanted with 0.3 at. % of helium (red trace). The data for the implanted material shows a slower decay, indicating a reduced thermal diffusivity, and a reduced SAW frequency, which suggests a lower stiffness.

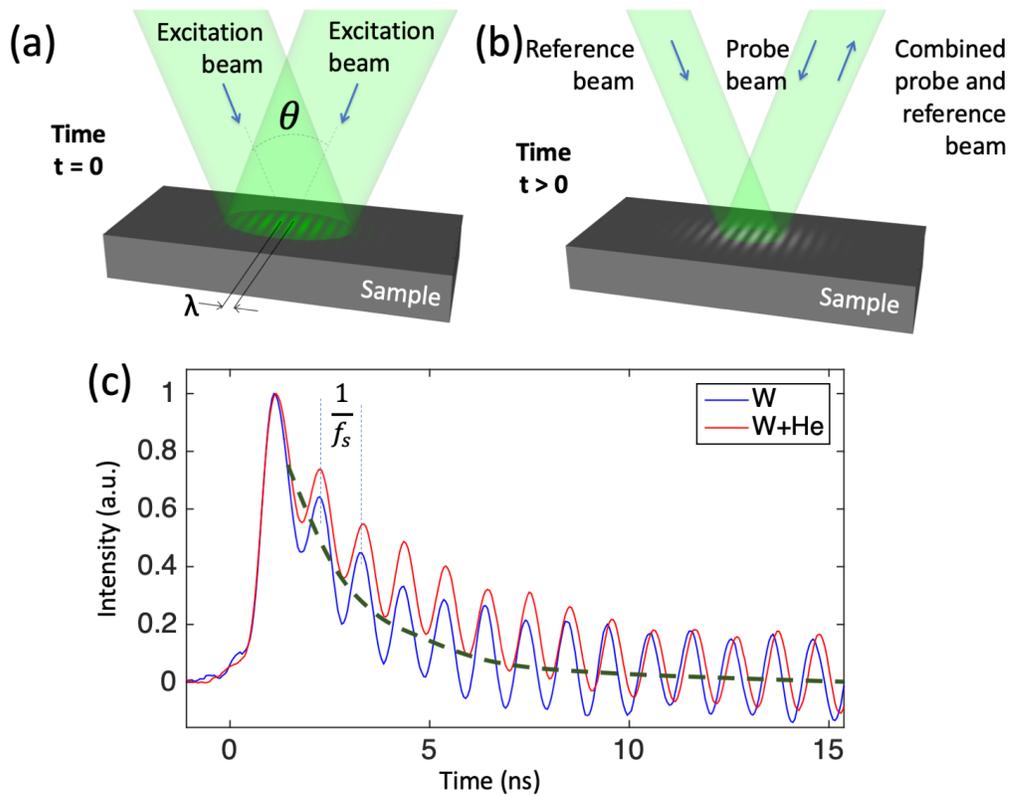

*Figure 2: (a) Generation of the transient grating by overlapping two short excitation beams on the sample. (b) Reading out of the transient grating by diffracting a probe beam from the grating in the sample, heterodyned with a reflected reference beam. (c) Transient grating signal from unimplanted tungsten (blue) and a helium-ion implanted region on the same sample (red). The oscillations of the signal reveal the SAW frequency, $f_s$. The underlying decaying background signal (green dashed as a guide to the eye) captures the decay of the temperature grating. Reproduced with permission from Hofmann et al. (35) under the [Creative Commons license 4.0](Creative Commons license 4.0).*

Quantitatively, the SAW velocity can be related to the material elastic constants of an elastically isotropic material using an approximate solution to the Rayleigh wave equation:

$$c_r = fL \approx (0.874 + 0.196\nu - 0.043\nu^2 - 0.055\nu^3)\sqrt{\frac{E}{2(1+\nu)\rho}}, \tag{2}$$

where *E* is the Young's modulus, *ρ* is the material density, and *v* is the Poisson's ratio, which holds for all values of $\nu \in [-1, 0.5]$ (*36*). For elastically anisotropic materials this relationship becomes considerably more complex due to the orientation dependence of elastic properties. This means that SAW velocity now depends on both out-of-plane crystal orientation and in-plane polarization of the SAW wave vector. Every et al. (*37*) and Favretto-Cristini et al. (*38*) reviewed different approaches for tackling this problem. Fortunately, several reliable computer codes for predicting direction-dependent SAW velocity have now been published (*39*, *40*). In addition, methods also exist for solving the inverse problem; namely, material properties may be extracted from measured SAW velocities (*41*).

The decay, or equilibration, of both the temperature and displacement gratings have been studied analytically in varying degrees of detail (*32*, *42*). Interestingly, the relaxation of the excited gratings in temperature and surface displacement are governed by different time dependencies. Following an impulse excitation and assuming an isotropic thermal diffusivity, *α*, the temperature grating, *T(t)*, varies as:

$$T(t) = \frac{A}{\sqrt{\alpha t}} \exp(-q^2 \alpha t), \tag{3}$$

whereas the surface displacement profile, *u(t)*, varies with time as:

$$u(t) = B\, erfc(q\sqrt{\alpha t}), \tag{4}$$

where *A* and *B* are proportionality constants and *q* is the grating wave vector $2\pi/L$ (*32*). In experiments, the temperature dynamics contribute to the overall measured response through the temperature dependence of the complex reflectivity of the surface. Johnson et al. (*32*) identified two broad classes of responses which may be recorded in TGS experiments, depending on the heterodyne phase used, i.e. the relative phase difference between the diffracted probe and reflected reference beams. For 'amplitude grating' responses, the reflectivity and, therefore, temperature dynamics may be isolated completely in the heterodyne response. In 'phase grating' responses, both reflectivity and displacement play a role and cannot be decoupled without knowledge of the precise heterodyne phase. When using TGS as a rapid NDE tool, the phase grating response is of greater interest as it encodes both thermal and elastic property information. Therefore, Dennett et al. recently proposed a robust new fitting approach that takes both relaxation dynamics into account, treating the ratio of their relative contributions as a free parameter (*43*).

An interesting question concerns the depth dependence of the measured signal. For both surface acoustic and thermal transport, the applied grating wavelength provides an easily tunable length-scale. For thermal grating decay, Käding (*42*) showed that the measurement is sensitive to thermal properties up to a depth of $L/\pi$, while a surface layer of thickness $L/2$ dominates the measured SAW signal (*44*). This ability to target different probing depths makes TGS measurements very attractive for selectively probing the properties of thin surface regions, or even depth profiling of properties (*45*). TGS is particularly useful for studying ion irradiation as it allows probing of the properties of the ion-irradiated layer alone, typically on the same order in thickness as the TGS probe depth, with little contribution from the unirradiated material beneath.

There are a few practical limitations to the extent of measurable material properties using TGS. The pump and probe lasers have fixed spot sizes (typically 50 to 500 μm) – if too small then the power density may be too high (the material under inspection is damaged by the excitation), if too large then the power density may be too low (the signal intensity is not large enough to measure). In addition, a sufficient number of projected laser fringes must exist within the pump spot to generate a sufficiently strong and monochromatic signal, placing an upper limit on the projected grating spacing. Meanwhile, as shorter wavelength SAWs have higher frequencies, the detection electronics – typically the avalanche photodiodes (APDs) and the oscilloscope – set a bandwidth limit for the fastest resolvable signal change. As of now, relatively inexpensive APDs with sufficient signal have an upper frequency cutoff around 1 GHz. This places a lower limit on the projected grating spacing for TGS – typically in the range of 2-3 µm for metals, 5-8 µm for ceramics. More fundamentally, the amplitude of the induced surface displacement scales with the wavelength of the imposed grating, leading to a reduction in recorded signal intensity as the wavelength is reduced.

A number of different geometries for TGS measurements have been proposed. Early implementation used beam-splitters and mirrors to split a short pump pulse and then generate two overlapping beams on the sample (*33*). However, this setup is challenging in practice as good phase stability between the pump pairs is needed, and it does not allow for straightforward, repeatable variation of the grating period. To overcome these limitations, Rogers et al. (*46*) suggested the boxcar geometry shown in Fig. 3(a), which is now used extensively. Here a volumetric diffraction grating, deemed a 'phase mask,' is used to split both pump and probe beams into many diffraction orders. The +1 and -1 orders are then isolated and overlapped on the sample using a 4f imaging system. Using shared optics for the pump and probe beams allows high relative phase stability. The grating period can be easily changed by changing the period of the phase mask. Indeed, using a tilted phase mask, a scheme for a setup that allows a continuously variable grating period was recently proposed (*47*). The boxcar geometry also allows straightforward setup of the heterodyne detection where the probe beam, diffracted by the transient grating in the sample, is heterodyned with a reflected reference beam (*31*). The phase difference between the probe and reference beams can be precisely tuned using an optical parallel inserted into the probe beam (labelled phase adjust in Fig. 3(a)). Until recently manual tuning of the phase was required after each measurement, leading to long measurement times of

~10 minutes per point depending on the material system and desired signal to noise ratio. Dennett et al. (*34*) proposed a dual heterodyne setup (Fig. 3(b)) that sidesteps this requirement, reducing scan times to a few seconds. This is key for enabling *in situ* measurements of evolving systems, as well as the mapping of larger areas.

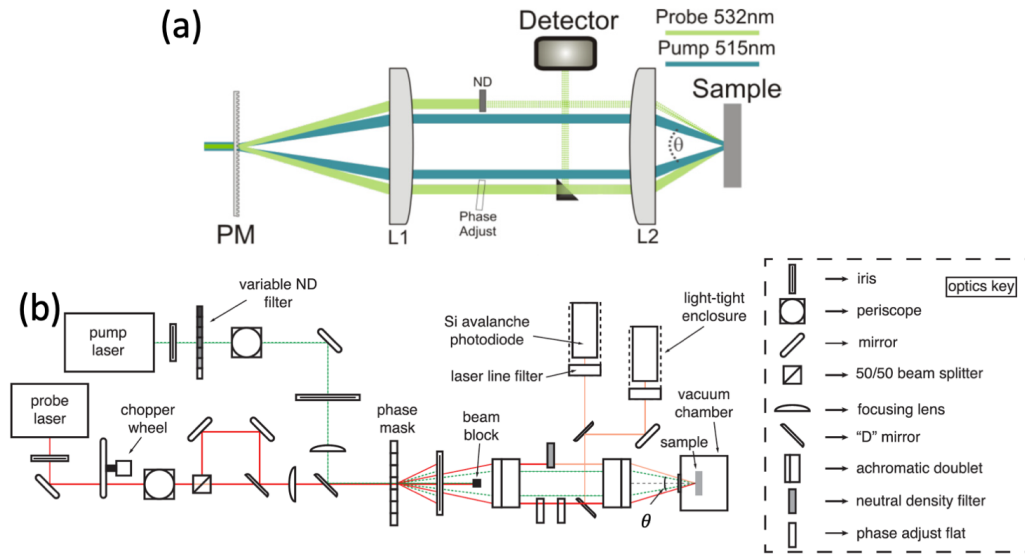

*Figure 3: Layout of a typical TGS experiment. (a) shows the boxcar geometry used to generate the transient grating. PM is a phase mask, L1 and L2 are achromatic doublets, and ND is a neutral density filter. Reprinted from Johnson et al.* (*32*), *with the permission of AIP Publishing. (b) shows the full layout for dual heterodyne detection TGS measurements as proposed by Dennett et al.* (*43*). *Reprinted from Dennett et al.* (*43*), *with the permission of AIP Publishing.*

**Applications:**

Applications of TGS range from direct probing of elastic and thermal properties, to using these properties to infer other information, for example to track the speed and extent of microstructural evolution under a number of driving forces. Here a few such illustrative use cases are presented.

*Probing elastic properties*

Directly measuring elastic properties is perhaps the most straightforward application of TGS, and the information is still surprisingly useful. The reason is that the elastic properties of even simple materials, e.g. pure Ni, have not yet been measured at moderate to high temperatures (*48*). In addition, it is important to chart out the ultimate sensitivity of the TGS technique when applied to a typical measurement in a reasonable amount of time, with signal collection times of tens of minutes. Dennett et al. (*48*) used TGS to measure the orientation-dependent SAW speed and elastic constants of pure Al and Cu single crystals and compared these with theoretical estimates (Fig. 4(a)). This study highlighted that TGS can measure SAW frequency with an absolute resolution of ~0.1% (the bandwidth of the SAW signal is typically 1% of the center frequency). This is a critical parameter, as it determines which types of changes should and should not be measurable using this technique.

Equally important to the sensitivity of a technique is its predictability, or the ability to pre-determine when a meaningful signal can be generated, as well as being able to understand signals measured in the lab. Dennett et al. (*48*) used large-scale molecular dynamics (MD) to create a scaled-down version of a TGS experiment in the computer, applying a 1D, spatially periodic, Gaussian pattern of heat to a box of atoms with periodic boundary conditions (Fig. 4(b)). The results agreed quite well with experimental measurements of the same materials, when properly scaled. As long as the MD simulation cell is larger than roughly 50nm in the grating dimension, and twice the grating dimension thick, the same acoustic wave propagation modes measured in experiment may be reliably detected in simulation. Although the wavelengths are much smaller in experiment, and therefore the frequencies much higher, the simulated acoustic wave *velocities* may be compared directly between simulation and experiment. Using this approach, the same relationship between vacancy concentration and Young's modulus as predicted by Dienes (*49*) and previously measured using contact ultrasonics (*50*, *51*) was obtained.

More recently, others have used TGS and similar SAW-generating techniques to direct probe elastic constants of simple and complex materials. Du and Zhao have used SAWs to measure the elastic constants of several pure metals (*39*) and a Ni-based superalloy (*52*), while Gasteau et al. have used SAWs to obtain single crystal elastic constants of polycrystalline steels (*53*). Indeed, using ultrasonic velocities to obtain crystal elastic constants is not new, as previous researchers have used other acoustic and/or ultrasonic techniques to measure the same (*54, 55*). However, as a non-contact technique, TGS lends itself particularly well to affixing a single crystal to a stage and rotating the sample to project the transient grating in different crystal directions. Particularly useful will be the ability to do so at high temperature, to map out some high temperature elastic properties, which remain unknown for most materials. An interesting question, yet to be answered, concerns the number of crystal rotations required to obtain a unique set of elastic constants for any given material. A definitive answer would be very useful for the ongoing and potential *in situ* and *in operando* applications of TGS outlined below.

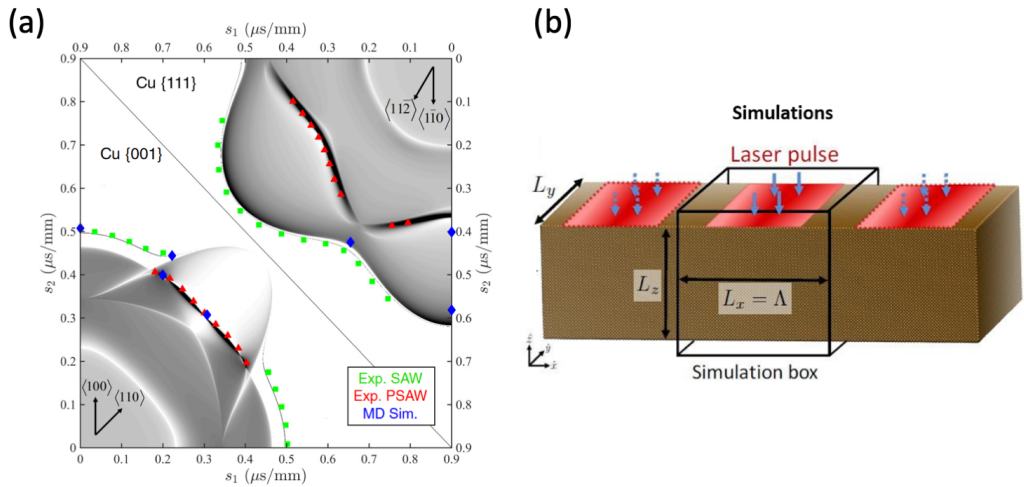

*Figure 4: Probing elastic properties by TGS. (a) Theoretical prediction (thin, black lines), experimental measurements (green and red for SAW and pseudo SAW respectively), and molecular dynamics predictions (blue) of SAW speeds on*

*single crystal copper* (*48*). *(b) Molecular dynamics simulation setup used to predict the orientation dependence of SAW speed in copper and aluminum* (*48*). *Reproduced with permission from Dennett et al.* (*48*) *under [Creative Commons license 4.0](#).*

*Probing material evolution via elastic properties*

The TGS technique becomes even more useful when detecting changes brought about by material processing, such as ion irradiation. A first demonstration of this from Hofmann et al. (*56*) considered the effects of helium ion implantation in tungsten, the main candidate material for plasma-facing armor in future fusion reactors (*57*). X-ray diffraction showed helium implantation-induced lattice swelling, and TGS a reduction in SAW velocity after implantation. These observations could be explained using density functional theory (DFT) calculations of the defects produced by helium ion implantation to predict the anticipated property changes. Lattice swelling and SAW speed reduction could then be interpreted in terms of the underlying population of helium-filled Frenkel point defects. Probing these defects by other means is tricky as TEM, for example, is not sufficiently sensitive to see them (*58*, *59*). Interestingly, the DFT calculations also predicted a very slight increase in elastic anisotropy from a Zener anisotropy ratio of 1.01 to 1.02 after ion implantation. This was confirmed by Duncan et al. (*60*) who measured an increase in Zener anisotropy ratio from 1.01 to 1.03 in a <110> oriented tungsten single crystals implanted with helium under the same conditions. Fig. 5(a) shows their measured SAW speed plotted as a function of direction in the (110) plane for the unimplanted and the implanted sample. Both the earlier observed SAW speed reduction, as well as the increase in elastic anisotropy upon implantation are clearly visible. This demonstrates the very subtle property changes that can be reliably quantified by TGS and linked back to the underlying defect microstructure.

Interestingly, TGS is not limited to detecting small lattice defects, but is also applicable to much larger scale defect structures. This is illustrated by a recent prototypical study in hastening the detection of radiation void swelling, a key degradation mechanism as described in the Introduction. In 2018, Dennett et al. (*61*) irradiated single crystal copper with 35 MeV self-ions, inducing void swelling as was clearly evidenced by TEM (see Fig. 5(c)). The onset of void swelling was linked directly to a drop in SAW velocity, backed up by TEM observations. However, this study still took nearly half a year to complete. Despite its clear utility, something far faster was required.

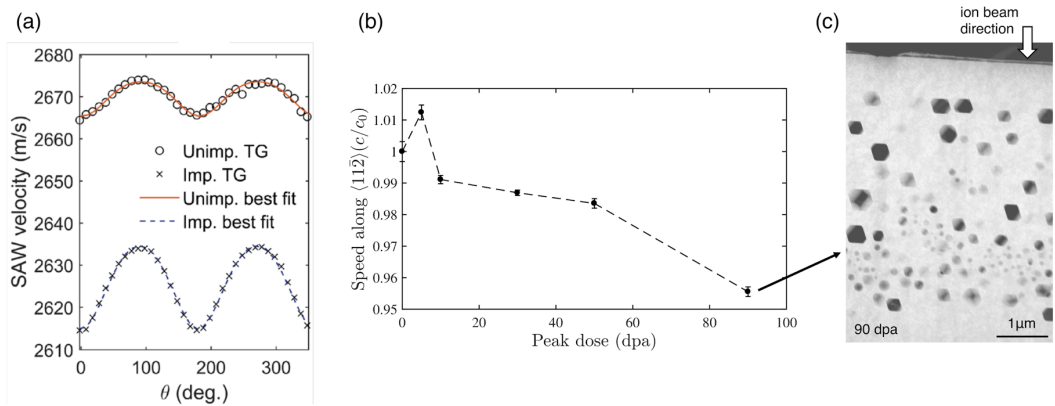

*Figure 5: Evolution of SAW velocity with low- and high-dose ion irradiation. (a) Orientation-dependent SAW velocity in unimplanted and helium-implanted regions of a <110> oriented tungsten single crystal. Reprinted from Duncan et al. (60), with the permission of AIP Publishing. (b) Measurement of void swelling in {111} oriented single crystal copper via SAW speed measurements along the ⟨11$\bar{2}$⟩ direction relative to that measured on the virgin sample. (c) HAADF STEM micrograph of the most highly exposed single crystal copper sample showing large amount of volumetric swelling. (b) and (c) are reprinted from Dennett et al. (61), with permission from Elsevier.*

An *in situ* TGS facility was therefore conceived and constructed (Fig. 6(a)) at the Sandia National Laboratory, pairing a dual heterodyne TGS facility with a 6 MV tandem accelerator (*62*). This allows for generation of heavy self-ions which can penetrate as deep as the TGS elastic and thermal probes – for the case of isotopes near the mass of copper 30-40 MeV ions are readily generated with typical ranges around 5 μm. TGS grating spacings of 2.5-8.5 μm are typical for many experiments, meaning that the depth of the ions and the depth of the material probed are well matched. Using this facility, an experiment similar to the single crystal Cu one was repeated *in situ* for single crystal Ni, producing a greatly *oversampled* curve of SAW speed vs. radiation dose (Fig. 6(b)). This reduced the time required for an experiment which would have taken months using traditional post-irradiation examination, and even weeks using *ex situ* TGS, to a single day. Both temperature and beam current are monitored and confirmed stable within reasonable limits, adding a measure of validity to the experiment. Performing all studies on the same sample also removes some of the variability associated with specimen preparation, accelerator performance, and other sample-to-sample effects.

For the *in situ* experiment in Fig. 6(b), the average dose between sequential data points, the dose resolution, is roughly 0.04 dpa. Traditional examination of materials undergoing this type of evolution results in data sampled with a resolution of up to 40 dpa between data points (Fig. 1) (*61*), resulting in a 1,000× dose resolution improvement when using *in situ* TGS. These new *in situ* experiments allow for far more precise determination of the dose to the onset of void swelling to be made, as well as the identification of new features not previously seen. For example, in Fig. 6(b) one can see an initial stiffening of the material, evidenced by a SAW speed increase. This stiffening is attributed to the interaction of dislocations and small defect clusters formed at low doses of irradiation (*61*). Under elastic loads, the native dislocation network will normally deform in a recoverable manner, with bowing segments pinned by forest dislocation interactions. Upon irradiation, small defect clusters serve as additional pinning sites for this native network and shorten the average pin segment length, which increases the observed elastic modulus (*63*, *64*). The saturation dose for this effect, as observed through TGS, matches well with *in situ* TEM observations of small defect cluster generation and saturation at low doses (*65*). As the applied dose increases, defects agglomerate to form larger-scale clusters, in this case voids, which then dominate the measured response.

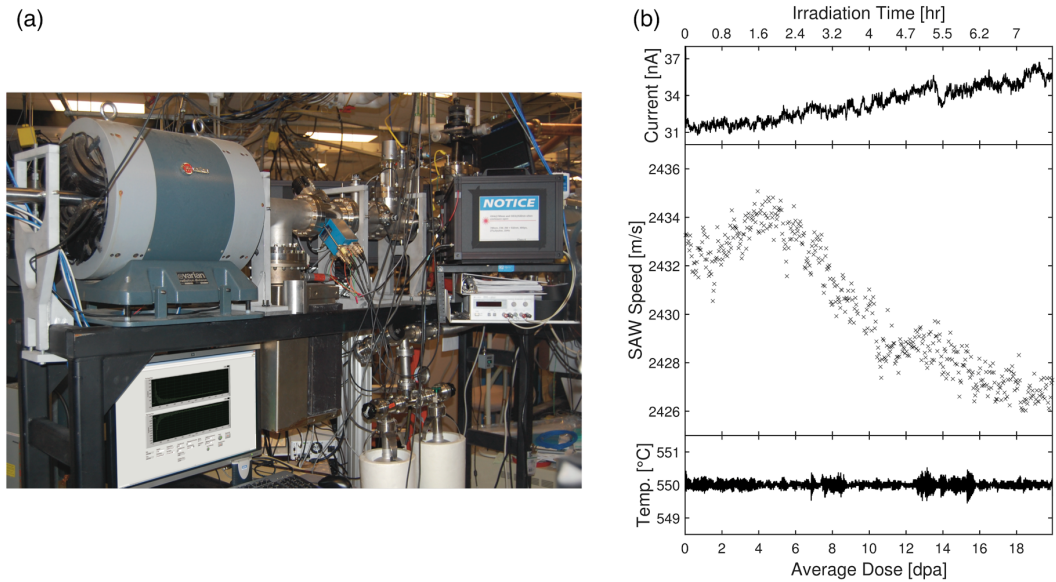

*Figure 6: In situ ion beam irradiation TGS (I³TGS). (a) Photograph of the I³TGS facility as installed at Sandia National Laboratories. (b) Applied 31 MeV Ni$^{5+}$ ion beam current, measured SAW speed, and temperature as a function of radiation dose in DPA in single crystal Ni, showing a drop in SAW speed corresponding to the onset of void swelling. Reprinted from Dennett et al. (62) with permission from Elsevier.*

### *Probing dynamics of micro-scale systems*

The ability of TGS to generate very well controlled, repeatable SAWs also provides an interesting opportunity for exploring micro- and nano-scale particle dynamics. For example, in granular media the nonlinear Hertzian contact between particles plays a key role in controlling properties. Colloidal crystals provide an attractive granular system where the collective response can potentially be tuned in a very controlled fashion, for example by varying particle composition, particle-particle and particle-substrate bonding, or particle size distribution. The first application of TGS to a 2D colloidal crystal of microspheres, deposited on top of a silica substrate coated with a thin aluminum film, was reported by Boechler et al. (66). Their experimental setup is shown in Fig. 7(a). By varying the TGS wavelength, and hence the SAW frequency, they could map out the dispersion response of the system (Fig. 7(b)). This reveals a number of interesting features: Longitudinal wave behavior in the sample (motion parallel to the substrate surface) is largely unaffected by presence of the microspheres. Rayleigh wave propagation, on the other hand, is strongly modified: Where the Rayleigh wave frequency approaches the natural frequency of the sphere-substrate contact resonance, an "avoided crossing" is seen due to hybridization of the two modes. This behavior can be reproduced using a surprisingly simple analytical mass-spring system model from which the microsphere-substrate contact stiffness can then be extracted.

A fine sampling of TGS wave vectors is desirable to map the dispersion curves in detail. Vega-Flick et al. (*47*) provided an elegant solution to this, using a tilted phase mask to generate a continuous variation of TGS excitation wavelength. Their approach allows much more detailed examination of the avoided crossing and phononic badgap. This is key for the design and testing of phononic materials with specific, engineered dispersion responses. The responses associated with other higher order dynamic modes, such as rocking of colloids, can also be examined using SAWs (*67, 68*).

An interesting question, particularly for the design of acoustic filters, concerns the attenuation of SAWs due to the presence of colloids. Eliason et al. (*69*) proposed a configuration with spatially separated excitation and probe beams to examine SAW attenuation by a strip of microspheres (Fig. 7(c)). Two counter-propagating SAWs are generated by the excitation, one of which then traverses the microsphere strip and is detected at the probe location using the heterodyne setup described above. Their results show a strong attenuation of SAWs at the contact resonance of the microspheres with a corresponding *1/e* attenuation length of less than two SAW wavelengths. Since microspheres can be easily deposited and removed, this opens up interesting new avenues for the manufacture of reconfigurable SAW devices with easily controllable structure and frequency selectivity, e.g. controlled by microsphere size. The Hertzian contact stiffness could also be modified, for example by functionalizing the surface of microspheres. Such devices could then serve as sensors where adsorption of species on the microsphere surface is detected based on the resulting modification of contact stiffness.

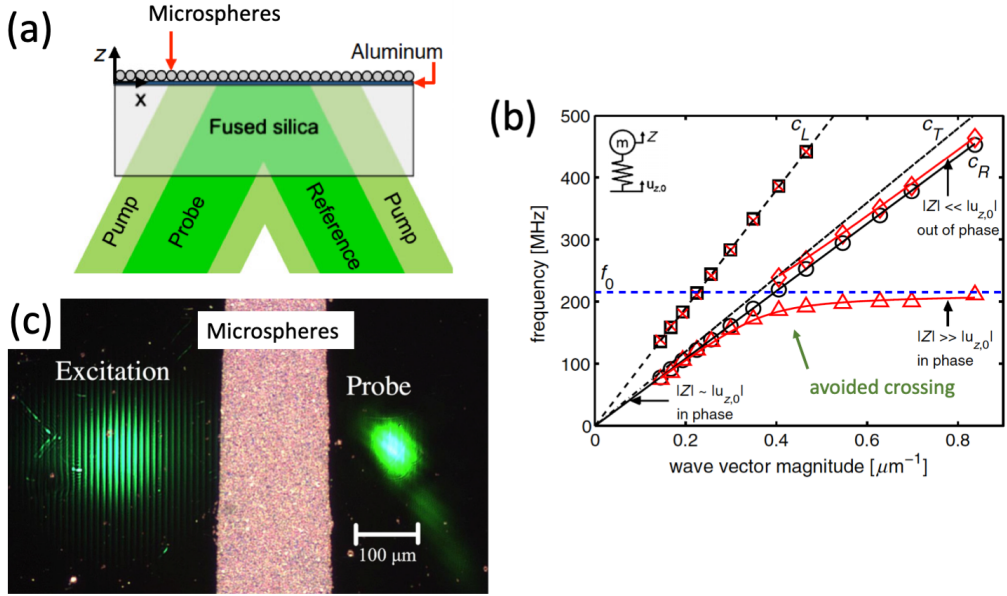

*Figure 7: Application of TGS to 2D colloidal crystals. (a) Schematic illustrating the TGS arrangement for measurement of the collective dynamics of microspheres deposited on an aluminum-coated fused-silica substrate. (b) Dispersion relations on spheres (red symbols) and off spheres (black symbols) for the arrangement in (a). Black lines show calculated longitudinal (subscript L), transverse (subscript T), and Rayleigh (subscript R) waves in fused silica. The blue dashed line shows the contact resonance frequency of the silica spheres. The solid red lines show the predicted dispersion for the substrate-microsphere system (66). (a) and (b) are reprinted with permission from Boechler et al. (66). Copyright (2019) by the American Physical Society. (c) Experimental configuration for measuring SAW attenuation by 2D colloidal crystals. Reprinted from Eliason et al. (69), with the permission of AIP Publishing.*

*Evolution of thermal transport properties*
Thermal conductivity is a key parameter for materials used in high heat load applications such as power generation, micro-electronics, or aerospace. A particularly useful example can be found in tungsten tiles used as plasma-facing armor in the divertor of future fusion reactors (*57*). Unexpected reduction of thermal conductivity caused by irradiation damage would lead to higher thermal stresses and larger surface temperatures, the key failure drivers of these components (*70, 71*). A first prototypical study using TGS to probe the effect of helium ion implantation damage on the thermal transport properties of tungsten was carried out by Hofmann et al. (*35*). The results (Fig. 8(a)) show that even a very small amount of implanted helium, 0.03 at. %, causes a reduction of room temperature thermal diffusivity by ~30%. Surprisingly, further increasing the implanted dose by a factor of ten only causes an additional ~20% drop in thermal diffusivity, suggesting the onset of a saturation regime. The question is how these changes can be understood in terms of the underlying defect population. In metals above the Debye temperature, thermal conductivity is dominated by electron-mediated transport (*72*). Implantation defects act as electron scattering sites, thereby reducing thermal conductivity. Surprisingly, whilst the effect of defects on phonon-mediated thermal transport is reasonably well understood and can be effectively modelled using atomistic simulations (*73, 74*), predicting the effect of complex crystal defects on electron-mediated transport remains challenging. Hofmann et al. (*35*) proposed the use of an atomistic model where the electronic scattering rate at atomic sites is assumed to be a function of the energy of each atom. A snapshot of this model is shown in Fig. 8(b) with atoms colored by electronic scattering rate. Vacancies appear as cages of eight atoms with increased scattering rate, whilst self-interstitial atoms, which in tungsten delocalize to form ½ <111> crowdions (*75*), appear as strings of atoms with increased scattering rate. Predicted profiles of thermal diffusivity for different defect concentrations agree surprisingly well with the experimentally measured curves (Fig. 8(a)). Whilst rather simplistic, this model provides a straightforward approach for predicting changes in thermal diffusivity due to complex crystal defect structures. The key challenge is knowing which defect population(s) should be simulated, since there are still monumental challenges associated with atomistic simulations of large levels of displacement damage (*76, 77*).

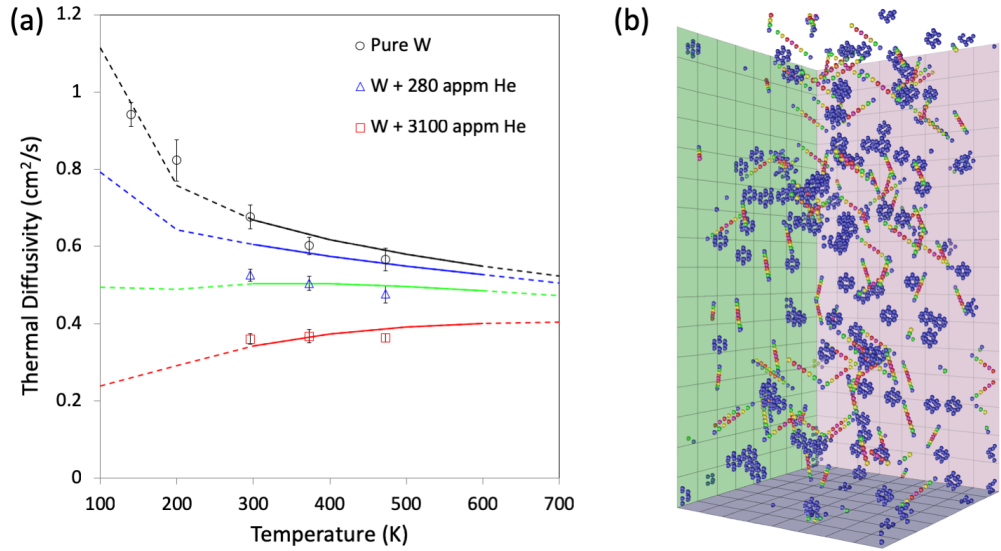

*Figure 8: Thermal diffusivity of helium ion implanted tungsten. (a) Experimentally measured thermal diffusivity (markers) of unimplanted tungsten, and tungsten implanted with 300 appm and 3000 appm of helium, plotted as a function of measurement temperature. Superimposed are predicted curves of thermal diffusivity for tungsten with Frenkel defect concentrations of 0 appm (black), 300 appm (blue), 900 appm (green), and 3000 appm (red). (b) Snapshot of the molecular statics model used to predict thermal diffusivity. Atoms are colored by electronic scattering rate from low (blue) to high (red). Reproduced with permission from Hofmann et al. (35) under [Creative Commons license 4.0](Creative Commons license 4.0).*

TGS measurements of thermal transport can also be used as a tool to pinpoint irradiation dose and to understand microstructural evolution. A recent example by Ferry et al. (*78*) explored the changes in thermal diffusivity upon the irradiation of single crystal niobium (Fig. 9), chosen for its high melting point and correspondingly low defect mobility at room temperature. Using TGS, the thermal diffusivity of Nb was found to drop by a factor of four, consistent with previous data that showed a rapid buildup of point defects at low doses (*79*). Upon further irradiation, the thermal diffusivity then recovered to half its pre-irradiation value. This can be explained by considering TEM data from Loomis and Gerber that showed initial generation of many tiny defects, followed by subsequent generation of larger defect clusters (*79*). This "microstructural cleanup," or clustering of radiation defects into fewer, larger agglomerates, is well known in radiation materials science, though using thermal diffusivity to detect it with TGS is a new development. Importantly TGS allows very rapid identification of these phenomena, without the need for slow microscopy sample preparation. This could be quite important in, for example, the design of a radiation damage-resistant material to pinpoint key stages of microstructural evolution. Given the similarities between the studies of Hofmann et al. and Ferry et al., it would be most useful to create an analytical, predictive model of TGS-measurable changes in thermal diffusivity upon irradiation, allowing for TGS-based determination of the dose to materials *a posteriori* in certain limited cases. This represents a first attempt at a potentially useful inverse problem – linking TGS-measurable material property changes to a dose in DPA, or more directly to material properties of direct interest.

The ability of TGS to selectively probe behavior of the thin ion-implanted layer is key for all of this. More bulk thermal analysis techniques like laser flash (*80–82*) would not have worked here, not to mention the small, irregularly-shaped specimen sizes being ill-suited to these more traditional techniques.

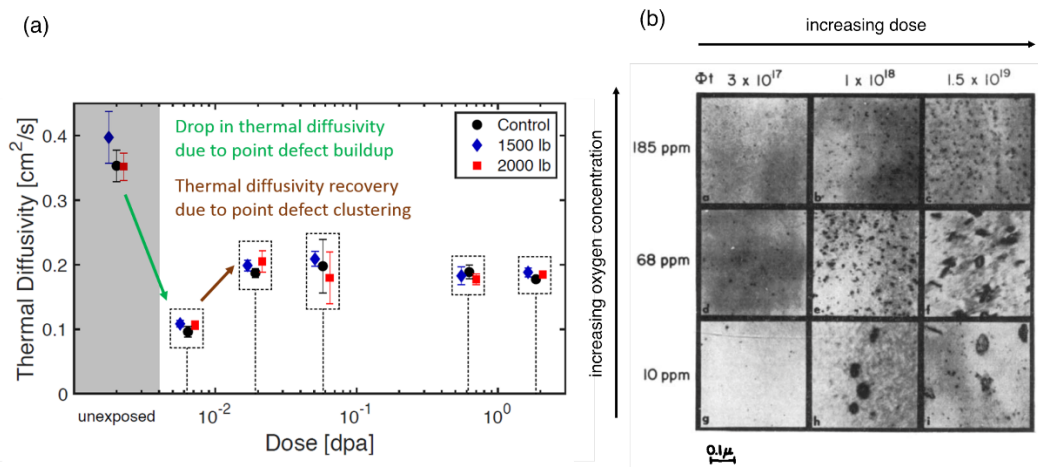

*Figure 9: Monitoring radiation-induced microstructural evolution via TGS in single crystal niobium* (*78*). *(a) A drop in thermal diffusivity signifies the generation of radiation point defects, while further irradiation causes fewer, larger clusters to form. (b) TEM results from a previous study* (*79*) *shows this*

*behavior occurring in single crystal Nb at similar radiation doses. Reprinted from Loomis and Gerber (79) with permission from Elsevier.*

### *Length-scale dependent thermal transport*

The characteristic length scale introduced by the transient grating wavelength is not only useful as a means of selecting the thickness of the surface layer probed. It also provides a convenient tool for examining the scale-dependence of thermal transport, i.e. the dependence of thermal conductivity on the distance over which thermal transport happens. In semiconductors, thermal conductivity is dominated by phonon-mediated transport (72). Despite a book value for phonon mean free path in silicon on the order 10s of nm (83), experimentally measured values of thermal diffusivity are known to be reduced when the length scale of the system in question is on the micron scale (84). The reason is that a large contribution to thermal diffusivity comes from low frequency phonons with mean free paths of up to several microns. TGS experiments have proven very useful in exploring this length scale dependence, making it possible to select portions of the phonon spectrum available to transport heat. As the grating period is reduced, phonons transition from a diffusive to a ballistic propagation regime, with phonons that have mean free path longer than the grating wavelength contributing little to thermal transport, thereby reducing the effective thermal diffusivity. The beauty of TGS is that different transport length-scales can easily be probed experimentally simply by selecting different applied grating wavelengths. Thus far TGS has been used to characterize the scale dependence of thermal diffusivity in a range of different semiconductors, such as silicon (84), germanium (43), SiGe (85), and GaAs (86).

### **Conclusion and Outlook:**

TGS is a powerful tool for the evaluation of key material properties ranging from elastic constants, to thin-layer thermal transport properties, to phonon attenuation. Importantly, since TGS only requires a reflective surface or a transparent sample, it is readily applicable to a whole host of different materials challenges and systems. By correlating TGS-measurable changes with microstructural evolution, TGS can further be envisioned as a non-destructive testing (NDT) tool for straightforward, non-destructive material health monitoring. For example, TGS could be used to inspect mission-critical components in light water reactors (LWRs), such as reactor pressure vessels or core barrels, to monitor material evolution and degradation. Using a miniaturized, portable TGS system this could be performed on in-service components, perhaps even without halting operation. The most exciting prospect for TGS though is to move beyond static measurements to *in situ* science. TGS allows for much faster probing than most traditional microscopy techniques, and since it does not require sample contact, it is perfectly suited to monitoring the *in situ* evolution of properties during complex material processing. Above we have discussed the current, rather limited number of *in situ* investigations, which focus on temperature and irradiation as the key driving forces. Many more applications could be envisaged, for example strain and electrical driving forces to monitor the *in situ* micro-scale behavior of actuation materials, perhaps even in miniaturized devices. TGS could also be used to monitor *in situ* chemical reactions, such as the growth of oxide films, making it an ideal tool for studying ultra-fast corrosion kinetics as long as the input laser energy does not affect the chemical reactions taking place. Indeed, provided the experimental geometries retain the necessary optical access for TGS monitoring, any combination of these driving forces may be applied simultaneously and synergistic effects studied. In all these scenarios, detailed knowledge of different microstructural evolution kinetics, ranging from precipitation/dissolution, to grain growth, to recrystallization via *in situ* TGS oversampling will allow for better tuning and validation of models and simulations. Such multi-property evolution maps may also be used to identify meaningful positions at which detailed characterization through other methods may be carried out to discover the underlying mechanisms controlling the evolution kinetics at hand. Developing and fully exploiting the *in situ* potential of TGS will be key for addressing the "characterization" bottleneck which is the time-limiting step for material innovation.

**Acknowledgments**

F.H. acknowledges support from the European Research Council (ERC) under Grant No. 714697. M.P.S. acknowledges support from the MIT-SUTD International Design Center (IDC) and the U.S. Nuclear Regulatory Commission (NRC) faculty development program under Grant No. NRC-HQ-84-15-G-0045 for funding the bulk of the TGS work in his group. C.A.D. acknowledges the US DOE NNSA Stewardship Science Graduate Fellowship under cooperative agreement No. DE-NA-0003864.

**Author biographies**

*Felix Hofmann*
Department of Engineering Science, University of Oxford, Parks road, Oxford, OX1 3PJ, UK
+44 1865 283448 (telephone)                              *felix.hofmann@eng.ox.ac.uk*

Felix Hofmann is an Associate Professor of Engineering Science at the University of Oxford, where the joined the faculty in 2013. His research concentrates on developing new X-ray and electron microscopy tools to probe the structure of atomic-scale crystal defects. These are combined with new micro-mechanics and laser-based techniques to clarify how defects modify mechanical and physical material properties. Most of his group's work is on materials for extreme environments in future nuclear and aerospace applications.

*Michael P. Short*
Department of Nuclear Science and Engineering, Massachusetts Institute of Technology, 77 Massachusetts Ave., Cambridge, MA 02139, USA
+1-617-347-7763 (telephone)                                      *hereiam@mit.edu*

Michael P. Short is the Class of 1942 Associate Professor of Nuclear Science and Engineering at the Massachusetts Institute of Technology. His research focuses on designing and diagnosing high performance materials for extreme service environments. Specific projects include the optical-based design of anti-fouling coatings for energy production, non-contact, non-destructive evaluation of materials degradation, and new ways of rapidly measuring corrosion and damage to materials.

*Cody A. Dennett*
Department of Nuclear Science and Engineering, Massachusetts Institute of Technology, 77 Massachusetts Ave., Cambridge, MA 02139, USA
+1-207-458-8399 (telephone)                                      *cdennett@mit.edu*

Cody A. Dennett is a Doctoral Candidate in the Department of Nuclear Science and Engineering at the Massachusetts Institute of Technology. His research focuses on the development of multi-modal, non-contact, and non-destructive materials evaluation techniques and the application of these methods for *in situ* monitoring of materials evolution during processing. Since 2016, Cody has been a US DOE National Nuclear Security Administration Stewardship Science Graduate Fellow.